\documentclass[twocolumn,showpacs,aps,prl,amsmath,amssymb]{revtex4}

\usepackage{bm,color}
\usepackage{graphicx}

\usepackage{subfigure}
\usepackage{psfrag}
\usepackage{epstopdf}
\usepackage{amsthm}

\renewcommand{\section}[1]{{\em #1}.---}
\newcommand{\letter}{paper}
\newtheorem*{theorem}{Lemma}
\newcommand{\erf}[1]{Eq.~(\ref{#1})}

\definecolor{nblue}{rgb}{0.2,0.2,0.7}
\definecolor{ngreen}{rgb}{0.2,0.6,0.2}
\definecolor{nred}{rgb}{0.8,0.2,0.2}
\definecolor{nblack}{rgb}{0,0,0}
\definecolor{mar}{rgb}{0.6,0.1,0.1}

\begin{document}
\title{Universality of the Heisenberg limit for estimates of random phase shifts}
\author{Michael J. W. Hall$^1$, Dominic W. Berry$^2$, Marcin Zwierz$^1$ and Howard M. Wiseman$^1$}
\affiliation{${}^1$Centre for Quantum Computation and Communication Techology (Australian Research Council), Centre for Quantum Dynamics, Griffith University, Brisbane, QLD 4111, Australia\\ ${}^2$Department of Physics and Astronomy, Macquarie University, Sydney, NSW 2109, Australia}

\begin{abstract}
The Heisenberg limit traditionally provides a lower bound on the phase uncertainty scaling as $1/\langle N\rangle$, where $\langle N\rangle$ is the mean number of photons in the probe.  However, this limit has a number of loopholes which potentially might be exploited, to achieve measurements with even greater accuracy.  Here we close these loopholes by proving a completely rigorous form of the Heisenberg limit for the average error over all phase shifts,  applicable to any estimate of a completely unknown phase shift.   Our result gives the first completely general, constraint-free and non-asymptotic statement of the Heisenberg limit. It holds for all phase estimation schemes, including multiple passes, nonlinear phase shifts, multimode probes, and arbitrary measurements.

\end{abstract}

\maketitle

Accurate estimation of optical phase is a highly important goal in quantum metrology.  For example, many optical measurement schemes, ranging from temperature sensing to gravitational wave detection,  rely on determining the phase shift induced by an environmental parameter \cite{phasesens}.  Further, phase-based optical communication relies on being able to decode information carried in the phase of an optical field \cite{phasecomm}.  It is therefore of fundamental theoretical and experimental interest to determine the maximum phase sensitivity that is possible when a given resource, such as the average photon number of a probe state, is available \cite{phaseres,multi,berry,nonlinear,exponential,zwierz,xiang,escher}.

A key concept in quantum phase estimation is the Heisenberg limit \cite{zwierz,heislim}, which  is generally understood as the asymptotic scaling relation
\begin{equation} \label{heislim}
\sigma(\hat{\Phi}) \gtrsim {k}/{\langle N\rangle}, \
\end{equation}
for any proportional measure of phase resolution, $\sigma(\hat{\Phi})$, 
where $\hat\Phi$ is the estimate of the true phase $\Phi$.   Here 
$k$ is a constant of order one, and $\langle N\rangle$ is the average of  the number operator which generates 
the phase shift, via the unitary operator $\exp(-i N \Phi)$. In the simple case of a linear, single-pass phase shift, 
$N$ is the total photon number (across all modes which experience the phase shift). 

It is well known that the Heisenberg limit is valid for canonical phase measurements on single-mode probe fields \cite{summy}, and hence for covariant phase estimates on such fields \cite{jmodopt}.  However, more generally,  the asymptotic bound in Eq.~(\ref{heislim}) is merely heuristic and open to challenge.  For example, general schemes for estimating phase shift parameters may include the use of multimode fields, multiple passes of probe states, nonlinear phase shifts, noncovariant phase measurements, and/or entangling joint measurements.  


Thus, there remain many loopholes which might potentially be exploited to achieve far greater phase measurement accuracies - for example, with asymptotic scalings better than $1/\langle N\rangle$ and/or arbitrarily small values of $k$ \cite{ani,zhang,rivas}.  It is, therefore, of considerable importance to determine whether this is possible.  

In this paper all such loopholes are closed  by giving a completely general constraint-free form of the Heisenberg limit, valid for all phase estimation schemes including multiple passes, nonlinear phase shifts, multimode probes, and arbitrary measurements.  Further, they are non-asymptotic, providing rigorous lower bounds on phase resolution valid for all values of $\langle N\rangle$.  

It will be shown  analytically that
\begin{equation} \label{varbound}
\delta \hat{\Phi} > {k_A}/
{{\langle N+1 \rangle}}  , 
\end{equation}
where $k_A := \sqrt{2\pi/e^3}  \approx {0.559}$ and $(\delta \hat{\Phi})^2$ denotes the average, over all applied phase shifts, of the mean square deviation of the phase estimate from the actual phase shift.  The asymptotic form of this bound immediately yields the Heisenberg limit in Eq.~(\ref{heislim}), with  $k = k_A$. 
Moreover, numerical work strongly supports the conjecture that $k_A$ in \erf{varbound} can be replaced by $ k_C \approx 1.376$, which is asymptotically optimal  as we show. 

It immediately follows from Eq.~(\ref{varbound}) that no 
scheme can do better, on average, than the Heisenberg limit.  It is crucial to the derivation of this bound that  the phase $\phi$ is {\em a priori} completely unknown, or random. That this is no impediment to entanglement-enhanced phase estimation was recently shown experimentally \cite{xiang}. 
Without such a condition, there is no universally valid Heisenberg limit of the form of \erf{varbound}.
In fact, there are some singular range restrictions for which {\em perfect} phase resolution is possible in principle, for a {\em finite} average photon number, as discussed further below.  

All of the results  in this \letter\ are proven using only the assumption that the eigenvalues of the generator $N$ are
nonnegative integers. In the optics context, this allows for the very general case of multimode (possibly entangled) states, with a phase shift generated by an optical nonlinearity \cite{nonlinear,zwierz}, 
and where the $k$th mode, with photon number operator $N_k$, passes through the phase shift $p_k$ times \cite{multi,berry}, i.e., $N = \sum_k p_k  (N_k)^q$, where $q$ and $p_k$ are integers.  Note that, in the case of nonlinear or multipass phase shifts, bounds scaling inversely with $\langle N \rangle$ do not  preclude a better scaling (e.g. inverse square) in terms of mean photon number, as many have noted \cite{nonlinear,zwierz}. 

Our results apply to any measurement whatsoever, including 
  adaptive measurements where each mode $k$ is measured separately, 
 but the result of measuring mode $k$ is used to alter the dynamics of mode $k'$ for $k'>k$ as in Refs.~\cite{multi,xiang}, 
 either by controlling $p_{k'}$ or by applying  (possibly nonlinear) auxiliary phase shifts on those modes.  This can be shown by extending and specializing the methods of Ref.~\cite{GLM06}. 

Giovanetti {\it et al}.\ have very recently obtained a related result, valid for estimation of any shift parameter, which for the case of phase estimation yields a bound on the average of the root mean square deviations for any two distinct applied phase shifts $\phi$ and $\phi'$:  
$\frac{1}{2}(\Delta_\phi \hat{\Phi} + \Delta_{\phi'}\hat{\Phi}) \geq \kappa(\lambda_{\phi\phi'})/\langle N\rangle$,
where  $\lambda_{\phi\phi'} := |\phi-\phi'|/(\Delta_\phi \hat{\Phi} + \Delta_{\phi'}\hat{\Phi})-1$. 
 The function $\kappa$ has a maximum of  $\approx 0.074$  at $\lambda_{\phi\phi'}\approx 4.7$.  This yields a 
Heisenberg limit, in an averaged sense, with scaling constant $k\approx 0.074$ \cite{glm},  valid  when the strong constraint $\lambda_{\phi\phi'}\approx 4.7$ is satisfied  or $\langle N\rangle$ is sufficiently large.  In contrast, the bound in Eq.~(\ref{varbound}) is entirely constraint-free and far stronger.

To proceed, consider a general phase estimation scenario in which a probe state, described by some density operator $\rho_0$ on an arbitrary Hilbert space,  undergoes a phase shift $\phi$ (the value of the random variable $\Phi$) to become
$\rho_\phi := e^{-iN\phi}\rho_0 e^{iN\phi}$. Here  $N$ is any operator with nonnegative integer eigenvalues.  A detection method $M$, described by some probability operator measure (POM) $\{ M_{\hat{\phi}}\}$, is then used to estimate $\phi$. That is, the probability density  of estimating a value $\hat{\phi}$, for the case of an actual phase shift $\phi$, is given by $p(\hat{\phi}|\phi) = {\rm tr}[M_{\hat{\phi}}\,\rho_\phi]$.

The overall performance of the estimate, for {\it a priori} unknown phase shifts, may be characterised by the concentration about zero  of the average probability density  
\begin{equation} \label{pbar}
\overline{p}(\theta)  =  \frac{1}{2\pi}\int_0^{2\pi}d\phi\, p(\theta+\phi|\phi) ,
\end{equation} 
of the random variable $\Theta:=\hat{\Phi}-\Phi \in [-\pi,\pi]$ 
 (see Fig.~1).   Note that for  a measure of concentration linear in $\overline{p}(\theta)$, such as $\langle \Theta^2\rangle$, this concentration is just the average of the individual concentrations of $p(\hat{\phi}|\phi)$ about $\phi$.    More generally, for a measure of concentration that is convex (concave), its value for $\overline{p}(\theta)$ is a lower (upper) bound on the average concentration. 

\begin{figure}[!t]
\centering

\includegraphics[width=0.40\textwidth]{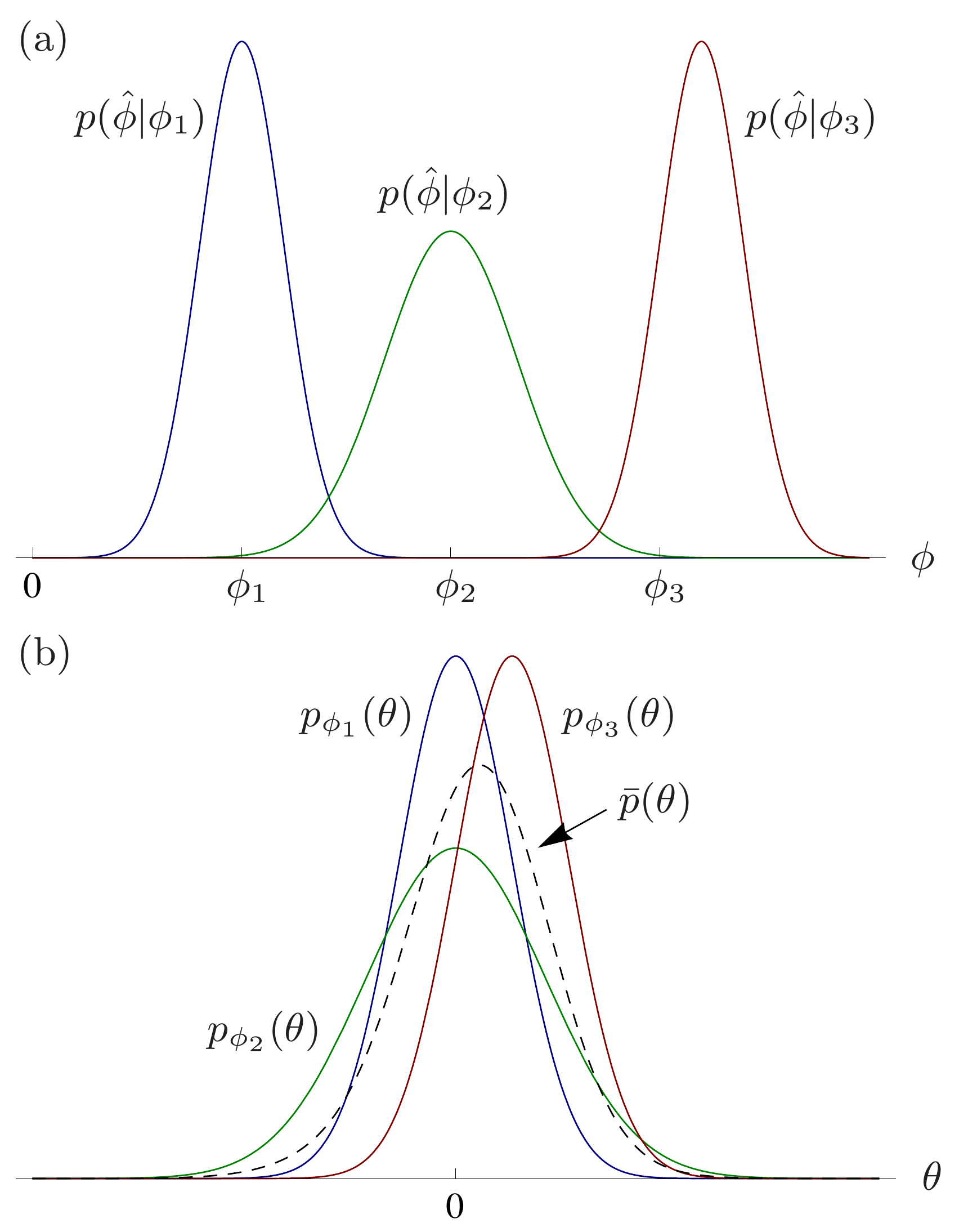}
\caption{(Color online) For a good estimate of a given phase shift $\phi$, the corresponding probability density $p(\hat{\phi}|\phi)$ must be well concentrated about the value $\hat{\phi}=\phi$. In the scenario depicted in 
(a),
the estimate is better for $\phi=\phi_1$ than for the cases $\phi=\phi_2$ and $\phi=\phi_3$ (in the latter case the distribution is offset from the actual phase shift $\phi_3$).  
The degrees of concentration can be compared more directly by translating them to the origin, as in 
(b). Thus, equivalently, a good estimate of $\phi$ corresponds to the distribution of $\theta=\hat{\phi}-\phi$,  i.e., $p_\phi(\theta) = p(\theta+\phi|\phi)$,
being well concentrated about $0$.  
It follows that how well the estimate performs overall, for randomly applied phase shifts, is characterised by the degree to which the {\it average} distribution,  $\overline{p}(\theta)$ in Eq.~(\ref{pbar}), is concentrated about $\theta=0$ (dashed curve).}

\label{fig:resolution}
\end{figure}

We  require  three known results. First,  the average phase distribution, $\overline{p}(\theta)$ in Eq.~(\ref{pbar})  can be  formally generated by a covariant phase estimate $\overline{M}$ for the same probe state \cite{holcov} (with POM $\{ e^{-iN\theta} \overline{M}_0 e^{iN\theta}\}$, where
$\overline{M}_0 := \frac{1}{2\pi} \int_0^{2\pi} d\phi \, e^{iN\phi}M_{\phi}e^{-iN\phi}$). 
Second, the phase distribution for a covariant phase estimate can be obtained via a canonical phase estimate, on a state with the same number distribution {of $N$} \cite{jmodopt,time}. 
Third, the canonical phase and number distributions of any {probe state} are identical to those of a corresponding single mode field \cite{berry,explicit}.


As a consequence of these three results,  the average phase distribution for {\it any} phase estimate is formally identical to the canonical phase distribution of some single mode field with the same number distribution. This yields the Lemma: 

\begin{theorem}   Any  bound on the concentration of the canonical phase distribution of a single mode field, under some constraint ${\cal C}$ on the photon number distribution, is also a bound on the concentration of the average phase distribution $\overline{p}(\theta)$ of an arbitrary phase estimate for any shift generator $N$ having nonnegative integer eigenvalues, providing that the probe state satisfies the same constraint ${\cal C}$ with respect to the distribution of {$N$}.  
\end{theorem}

Note that $N$ may be any generator with nonnegative integer eigenvalues, and its spectrum may be a strict subset of the nonnegative integers (as in the case of optical nonlinearities).  
It is clear that all bounds will still hold under such a constraint on the spectrum, though it means that there may be tighter bounds.  

We now use the above Lemma   to obtain the bound in Eq.~(\ref{varbound}).
The average of the mean square deviation of an estimated phase shift from the actual phase shift, over all possible phase shifts $\phi$, can be written as
\begin{equation} \label{var}
(\delta \hat{\Phi})^2 :=  \langle\Theta^2\rangle= \int_{-\pi}^\pi d\theta\, \theta^2\,\overline{p}(\theta)=\frac{1}{2\pi} \int_0^{2\pi} d\phi\, {\rm Var}_\phi  \hat{\Phi}  ,
\end{equation} 
where ${\rm Var}_\phi \hat{\Phi} :=\int_{\phi -\pi}^{\phi+\pi} d\hat{\phi} \, (\hat{\phi}-\phi)^2\,p(\hat{\phi}|\phi)$ \cite{jmodopt}. The entropic uncertainty relation for the canonical phase and photon number observables of single mode fields \cite{jmodopt,bbm} immediately yields, via the Lemma, $H({\Theta}) + H(N) \geq \log_e 2\pi$ for the entropies of  $\Theta$ and $N$. Further, formally extending $\overline{p}(\theta)$ from $[-\pi,\pi]$ to $(-\infty,\infty)$, we obtain $H(\Theta) <\frac{1}{2}\log_e[2\pi e(\delta \hat{\Phi})^2]$ \cite{Cover} (where the inequality is strict since it is only saturated by Gaussian distributions over the whole real line).  
Because entropy is maximised for thermal distributions, $H(N)\leq \log_e \langle N +1\rangle + \langle N\rangle\log_e (1+1/\langle N\rangle)$. Noting the last term is always less than unity, these inequalities combine to yield
\begin{equation} \label{chain}
(\delta \hat{\Phi})^2 >  \frac{2\pi}{e} e^{-2H(N)} > \frac{(2\pi/e^3)}{\langle N +1\rangle^2} ,
\end{equation}
and, hence, Eq.~(\ref{varbound}) as desired.  

Thus we have shown that the Heisenberg limit holds in general.
There is no way to take advantage of special measurement schemes in order to obtain measurements with scaling better than $1/\langle N \rangle$, provided the phase is initially unknown.
By using $\langle N+1\rangle$ in the denominator, rather than $\langle N\rangle$, we provide a bound that holds for all $\langle N\rangle$; it is clear that there cannot be bound of the form $k/\langle N\rangle$ which is valid for all $\langle N\rangle$, because  the phase error cannot be larger than $\pi$. 
Note that for the special case of $m$ identical probe states, each with mean number $\langle N_1\rangle$, our lower bound evaluates to $k_A/\langle mN_1+1\rangle$.

A secondary question is whether, and by how much, the scaling constant in Eq.~(\ref{varbound}) can be improved.  Here we report strong numerical evidence for the conjecture that
\begin{equation} \label{conj}
\delta \hat{\Phi} > { k_C }/{\langle N+1\rangle }, 
\end{equation}
 with $k_C  :=2(-z_A/3)^{3/2}\approx 1.37608$, is the best possible lower bound of this form, where $z_A$   denotes  the first (negative) zero of the Airy function.  In particular, the minimum possible value of $\delta\hat{\Phi}$ for any canonical phase estimate on a single mode field, and therefore for any general phase estimate via the Lemma, may be numerically calculated as a function of $\langle N\rangle$ via standard techniques.

In particular, one can use the method of Lagrange multipliers, as described in Ref.\ \cite{berrythesis}.
In this case, rather than maximising $\langle \cos\theta\rangle$, one wishes to minimise $\langle \theta^2\rangle$.
Because $\langle \theta^2\rangle$ is symmetric and considered over the region $[-\pi,\pi]$, it can be expanded as a cosine series.
Then the Lagrange multiplier problem may be formulated as an eigenvalue problem, and solved numerically.
The resulting minimum values of $\delta\hat{\Phi}$ are plotted as a product with $\langle N+1\rangle$ in the upper curve in Fig.~2.
It is seen that the value of the product asymptotically approaches a minimum value from above that is numerically indistinguisable from $k_C$.

\begin{figure}[!t]
\centering
\includegraphics[width=0.45\textwidth]{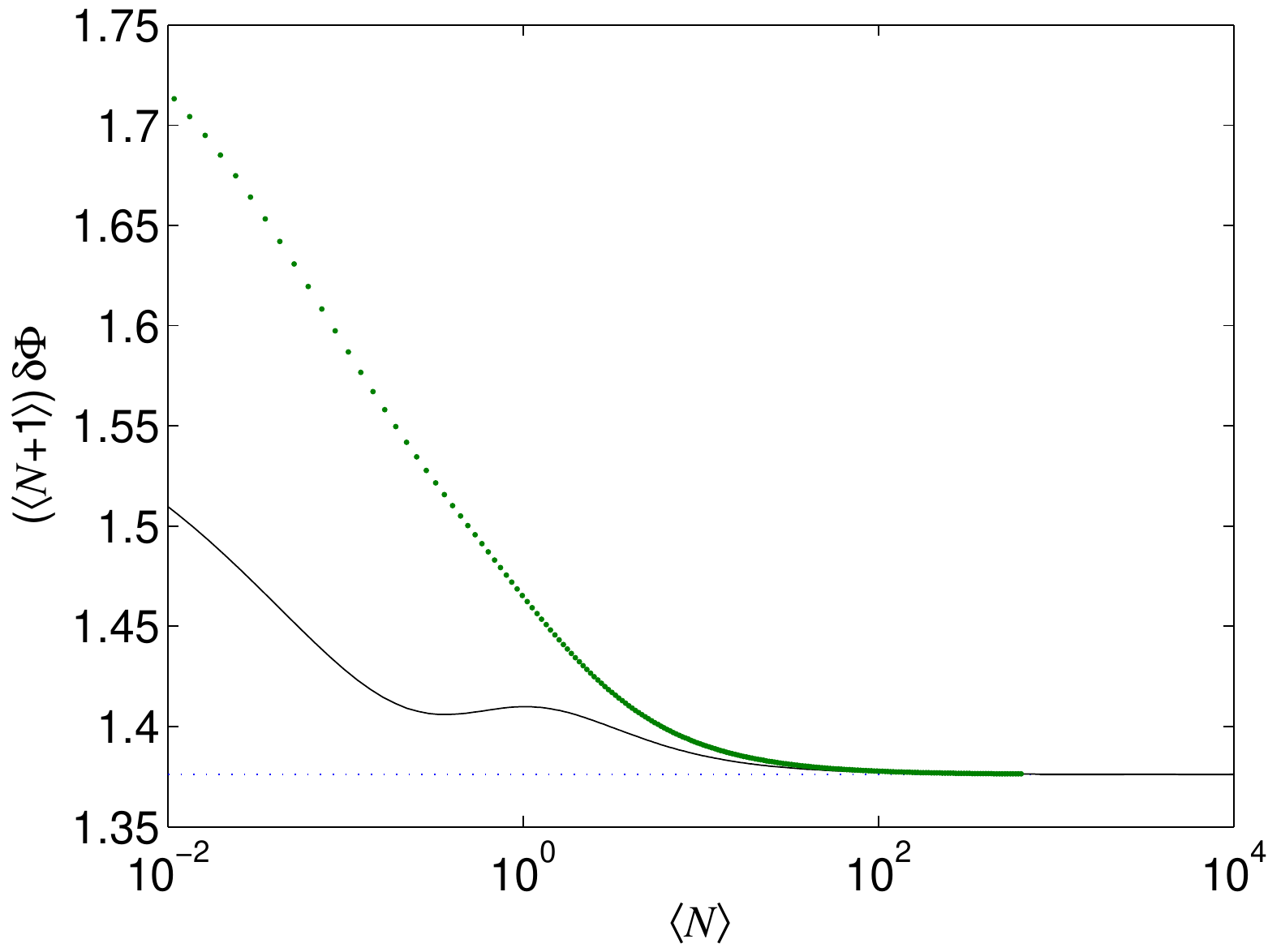}
\caption{(Color online) The minimum possible value of $\langle N+1\rangle \delta \hat{\Phi}$, plotted as a function of $\langle N\rangle$ (upper curve).  The aymptotic value is numerically indistinguishable from $ k_C \approx 1.37608$ in Eq.~(\ref{conj}) (lower dotted curve).  As per the text, the asymptotic value cannot be any less than $ k_C$.  The intermediate curve replaces $\theta^2$ in Eq.~(\ref{var}) by the lower bound $\frac{5}{2}-\frac{8}{3}\cos\theta +\frac{1}{6}\cos 2\theta$, for comparison purposes.  This can be computed efficiently for larger values of $\langle N\rangle$, and shows the same asymptotic behaviour. }
\label{fig:numerical}
\end{figure}

Moreover, because  $\frac{5}{2}-\frac{8}{3}\cos\theta +\frac{1}{6}\cos 2\theta$ is no greater than $\theta^2$, the minimum expectation value of the first function  also places a lower bound on the minimum phase variance.
This optimisation is computationally easier, and can be performed for larger values of $\langle N \rangle$, because the eigenvalue problem obtained is sparse.
The minimum values obtained are shown as the intermediate curve in Fig.~2.
Again these results do not drop below $k_C$, providing further evidence for the inequality \eqref{conj}.

The constant  $k_C$ can be obtained from the properties of the Holevo variance,
$(\delta_H \hat{\Phi})^2 := |\langle e^{i\Theta}\rangle|^{-2} -1$ \cite{holcov,WisKil97}.  For a canonical phase measurement on a single mode field, this has a tight asymptotic bound $\delta_H\hat{\Phi}\gtrsim k_C /\langle N\rangle$, valid to first order in $\langle N\rangle^{-1}$ \cite{berrythesis,bandilla}.  The inequality chain $|\langle e^{i\Theta}\rangle|^2 \geq \langle\cos\Theta\rangle^2 \geq \langle 1-\Theta^2/2\rangle^2$ then yields
\[ (\delta_H \hat{\Phi})^2 \leq \left[ 1-\frac{(\delta \hat{\Phi})^2}{2}\right]^{-2}-1 = \sum_{m\geq 1}(m+1)\left[ \frac{(\delta\hat{\Phi})^2}{2}\right]^{m} \]
for general estimates, via the Lemma. In the asymptotic limit only the $m=1$ term contributes, to lowest order, and hence $\delta \hat{\Phi}\gtrsim  k_C/\langle N\rangle$. 
Thus, the constant $k_C$ in the conjectured bound of Eq.~(\ref{conj}) is asymptotically optimal. 

 Bounds on other direct measures of overall phase resolution can be similarly obtained.  
For example,  the  `ensemble length'
$L(\hat{\Phi}) := e^{H(\Theta)}$,  which quantifies the effective support length of $\overline{p}(\theta)$ 
\cite{entvol},  has the analytic lower bounds
$L(\hat{\Phi}) \geq 2\pi e^{-H(N)} > {(2\pi/e)}/{\langle N+1\rangle}$,
analogous to Eq.~(\ref{chain}), using the  same entropic uncertainty relation as before.  
The bound is  stronger than, and implies, the bound in Eq.~(\ref{varbound}). 
 The Lemma and general methods above can also be used to obtain similar bounds on overall phase resolution for other constraints on $N$, such as having a fixed maximum 
value $n_{\rm max}$ \cite{future}.

The above results establish the validity of the Heisenberg limit, as a strong constraint on the overall performance of general phase detection schemes.  Indeed, Eq.~(\ref{varbound}) establishes a non-asymptotic analytic and constraint-free lower bound for average phase resolution. Furthermore, strong numerical support has been given for the conjecture that the  numerator in Eq.~(\ref{varbound}) can be replaced by the asymptotically optimal scaling constant $ k_C$, as per Eq.~(\ref{conj}).  All results immediately generalise to bounds on the time resolution of any periodic system, with Hamiltonian $H=E_0+\hbar\omega N$,  via the replacement  $\hat{\Phi}\rightarrow \omega \hat{T}$.  
{{They can also be generalised to shift generators having arbitrarily  negative integer eigenvalues \cite{future,neg}.}

It follows that it is only possible to go beyond the Heisenberg limit when there is some restriction on the applied phase shift $\phi$, i.e., when prior information about $\phi$ is available.
In such cases one should not average over $\phi$ with equal weighting, but instead should use a weighting corresponding to the prior probability.
Indeed, there is the well known case where the phase shift is limited to values in the finite set $\{ \phi_k := 2\pi k/K\}$, where $k=0, 1, \dots K-1$ for some integer $K\geq 1$.
Choosing the single mode probe state $\rho_0=|\psi\rangle\langle\psi|$, with $|\psi\rangle := K^{-1/2}\sum_{n=0}^{K-1} |n\rangle$, and a phase estimate corresponding to measurement of the Hermitian operator ${\Phi}_K:= \sum_{k=0}^{K-1} \theta_k e^{-iN\phi_k}|\psi\rangle\langle \psi|e^{iN\phi_k}$, then the phase-shifted states $\rho_{\phi_k}$ are distinct eigenstates of ${\Phi}_K$.
Hence, the phase shifts can be resolved without error, even though the average photon number is finite.

Another case is that where the error in the phase estimate is smaller than the Heisenberg limit only for some specific value of the phase.
It might be imagined that one could take advantage of this measurement (without prior information about the phase) by using initial measurements to determine an initial estimate of the phase.
Our results show that no such scheme can do better than the Heisenberg limit  overall:
when one takes into account the resources needed to initially estimate the phase, the complete measurement cannot beat the Heisenberg limit. 

It would be of interest to consider the case where such prior information is characterised by the properties of some prior probability density, $q(\phi)$,  so that the factor $1/(2\pi)$  in Eq.~(\ref{pbar}) is replaced by integration over $q(\phi)$. 
 This case will be addressed in future work.

MJWH, MZ, and HMW are supported by the ARC Centre of Excellence CE110001027. DWB is funded by an ARC Future Fellowship (FT100100761).


\begin{thebibliography}{99}
\bibitem{explicit} To give an explicit construction,
let $\{|n,d_n\rangle\}$ denote a complete basis set of eigenstates of  the  generator $N$, where $d_n=1,2,\dots,D(n)$ labels the degeneracies of eigenvalue $n$.  For any probe state $\rho$ define the single mode state $\rho_s$ via  $\langle m|\rho_s|n\rangle :=\sum_{d\leq D(m),D(n)} \langle m,d|\rho|n,d\rangle$ for $m,n\in P$ and vanishing otherwise.   The number distribution of $\rho$ is then identical to the photon number distribution of $\rho_s$ and, similarly, the canonical phase distribution for $\rho$, generated by the POM $\{ e^{-iN\phi}C_0e^{iN\phi}\}$ with $C_0 = \frac{1}{2\pi} \sum_d \sum_{\{n,n':d\leq D(n),D(n')\} } |n,d\rangle \langle n',d|$ \cite{time}, is identical to the canonical phase distribution of $\rho_s$.
\bibitem{neg} {For such a generator $J$ (e.g., $J=N_1-N_2$), the single mode field of the Lemma is replaced by a rigid rotator on a circle.  The first inequality in Eq. (\ref{chain}) remains valid with $N$ replaced by $J$, leading to, e.g., a bound as per Eq.~(\ref{varbound}) with $N$ replaced by $2|J|$ \cite{future}.}
\end{thebibliography}
\end{document}